\documentclass{epl}

	\title{Continuous optical loading of a Bose-Einstein condensate in the Thomas-Fermi regime}

\author{F. Floegel, L. Santos and M. Lewenstein}

\institute{Institut f\"ur Theoretische Physik, Universit\"at Hannover,
 D-30167 Hannover,Germany}


\newcommand{\hc}{\mathrm{h}.\mathrm{c}.}

\begin{document}

\maketitle

\begin{abstract}
We discuss the optical loading of a Bose-Einstein condensate in the 
Thomas-Fermi regime. The condensate is loaded via spontaneous emission 
from a reservoir of excited-state atoms. By means of a master equation 
formalism, we discuss the modification of the condensate temperature 
during the loading. 
We identify the threshold temperature, $T_{th}$, above (below) which the
loading process leads to cooling (heating), respectively.
The consequences of our analysis for the continuous loading of an atom laser 
are discussed.
\end{abstract} 

The experimental realization of a Bose-Einstein condensate (BEC)
has aroused an enormous interest during the recent years \cite{Varenna}. 
Among the disparate variety of topics related with the BEC physics, 
the possibility of accomplishing a coherent source of atoms, or atom laser, 
has attracted a large attention \cite{AtomLaser}. As a coherent source of matter waves, 
the atom laser is expected to lead to new applications in atom optics, 
and its relevance is often compared to the impact of optical lasers in
standard optics. The atom-laser has been produced experimentally after outcoupling 
atoms from a BEC either by employing radio-frequency fields \cite{rf,Bloch}, or Raman techniques \cite{raman}.
The coherent character of the source 
has been demonstrated in a series of interferometric experiments \cite{Bloch,MIT2,Kasevich}.
The availability of continuous-wave (cw) atom 
lasers would open the way to "high power" and precision applications.
Several groups have developed various techniques to outcouple atoms in a (quasi-) continuous way 
\cite{raman,Bloch}. However, the continuous 
outcoupling represents just a half way towards a cw atom laser. 
Without a continuous refilling of the BEC, the atom
laser output lasts only as long as some atoms in  the BEC are kept. 
In addition, the continuous refilling of the condensate could be employed to repair 
a condensate against inelastic losses, and in this way it could allow to 
analyze the BEC physics at larger densities, at much longer time scales. Recently, 
the first steps toward this direction has been achieved in an experiment performed 
at MIT \cite{MITrefill}.

Two different physical mechanisms could provide a continuous
pumping into a BEC. On one hand, collisional mechanisms \cite{Castin}, 
in which two non-condensed reservoir atoms collide, 
one being pumped into the BEC, 
whereas the other carries most of the energy and is evaporated.
Several proposals have been suggested in this direction. 
Ref.\ \cite{Castin} analyzes the continuous evaporative cooling 
of an atomic beam directed along a long magnetic guide, at the end of which 
the atoms are predicted to enter the degenerate regime. This method 
demands a large flux of slow atoms in order to achieve a 
sufficiently large collisional rate which guarantees the required efficiency of the 
evaporative cooling in a finite guide.
Recently, Roos {\it et al.} \cite{roos} have analyzed a scheme in which particles from an incident beam are trapped 
in a potential well when colliding with particles already present in the well. 
For a cigar-shape potential, it has been shown that a resonance can occur when the transverse evaporation threshold 
coincides with the energy of the incident particles, leading to a large increase in phase space density 
with respect to the incident beam. 
We should also refer at this point to a recent experiment \cite{MITrefill}, where, by means of 
an optical tweezer, a BEC is merged with a pre-existing one. 
A second way of continuous loading of a BEC is provided by the 
spontaneous emission from a reservoir of excited-state atoms \cite{Martin,Pfau,Pfau2}. 
If this reservoir could be filled
in a (quasi-) continuous way  by laser cooling techniques, one
would benefit from the large cooling efficiency of laser cooling
compared to evaporative cooling, allowing for a considerable
increase in atomic flux produced by an atom laser. 
In order to achieve the continuous optical loading of a BEC,  
it is crucial to avoid the heating introduced by the photon reabsorption \cite {Dalibard}. 
Fortunately, it has been shown that in the 
so-called Festina Lente regime, in which the spontaneous emission rate, $\gamma$, is 
smaller than the frequency of the harmonic trapping, $\omega$, 
the heating due to reabsorption is practically suppressed  \cite{Festina}. 
It has been also recently reported that, under certain conditions, the reabsorption can be also largely suppressed 
in atoms with an accessible three level $\Lambda$ scheme \cite{Pfau}. 
Additionally, quantum interference effects can lead to positive (cooling) 
effects of the reabsorption  in the so-called BAR regime \cite{BAR,BAR2}.\\
\indent
The continuous optical loading of a BEC have been previously analyzed in 
ref.\ \cite{Pfau}, in which 
the BEC was considered in the weak condensation regime, where the mean interaction energy is 
smaller than the separation between the harmonic levels. In that case, the 
loading dynamics can be split from the collisional physics in the corresponding quantum Master 
equation (QME), and both phenomena can be simultaneously simulated via Monte Carlo methods
\cite{Lascoll,Pfau}. In particular in ref.\ \cite{Pfau}
, it was shown that 
a BEC can be first created in an initially empty trap by means of optical loading, and then 
maintained against losses by properly fixing the pumping rate.
However, the weakly-interacting regime sets important restrictions to the atomic density. 
In the present paper, we extend our analysis to 
arbitrary regimes, and in particular to the Thomas-Fermi regime,
in which the mean interaction energy 
is much larger than the separation between the trap levels. In that case, the BEC wavefunction 
and the excitations depend on the particle number, and as consequence the 
wavefunctions vary dynamically during the pumping process, complicating the analysis of the problem.
In this paper we develop the necessary theory to 
analyze the problem. In particular, we study the modification of the condensate temperature 
during the pumping process.  
We show that there is a threshold temperature for a
given number of trapped particles above (below) which the system is cooled (heated), and show that this effect could be employed to control the condensate 
temperature.\\
\indent
We consider a sample of cold atoms of mass $m$
with an accessible electronic two-level system, formed by the 
ground state $|g\rangle$ and an excited state $|e\rangle$. Both $|g\rangle$ and $|e\rangle$ atoms 
are in a harmonic trap, which for simplicity 
is considered isotropic, with frequencies $\omega_g$ and $\omega_e$, respectively. 
The $|g\rangle$ atoms are Bose condensed, {\it i.e.} 
with a temperature $T\ll T_c$, where $T_c$ is the critical temperature 
for onset of the condensation. The formation of the condensate using spontaneous emission 
was previously discussed in ref.\ \cite{Pfau}. The $|e\rangle$ atoms, which eventually decay spontaneously into the 
$|g\rangle$ state, are considered as thermally distributed.\\
\indent 
The atoms in the $|g\rangle$ trap are described by the corresponding field operator 
$\hat{\Psi}_g(\vec{r})$. For sufficiently low $T$, when the number of condensed atoms in the $|g\rangle$ trap, $N_0$, 
is comparable to the total number of atoms, $N$, one can employ the Bogoliubov approximation and  
substitute $\hat{\Psi}_g$ by a $c$-number, $\psi_{0}$, the BEC wave function, whose  
stationary value is provided by the time-independent Gross-Pitaevskii equation (GPE)
\begin{equation}
\mathcal{H}_{GP}\psi_{0}(\vec{r})=\left( -\frac{\hbar ^{2}}{2m}\nabla^2 +\frac{m}{2}\omega_g^2 r^2+
U_{gg}N|\psi_{0} |^{2}-\mu \right) \psi_{0} (\vec{r})=0,
\label{GPE}
\end{equation}
where $\mu$ is the chemical potential, and $U_{gg}=4\pi\hbar^2a_{gg}/m$ 
is the coupling constant for the collisions between $|g\rangle$ atoms, with 
$a_{gg}$ the scattering length. In eq.\ (\ref{GPE}) we have neglected the $e$-$g$ collisions, 
since the number of atoms in the $|e\rangle$ state is assumed very small.

The excitation spectrum is obtained after linearizing $\hat{\Psi}_g$ around the ground-state solution, 
$\hat{\Psi}_g(\vec{r})=\psi_{0}(\vec{r})+\delta \hat{\psi }(\vec{r})$.
The perturbation $\delta\hat{\psi}$ is then expanded in the standard form \cite{Stringari}
$\delta \hat{\psi }(\vec{r})=u_{\vec{n}}^{*}(\vec{r})\tilde{g}_{\vec{n}}
-v_{\vec{n}}(\vec{r})\tilde{g}^{\dag }_{\vec{n}}$,
where $ \tilde{g}_{\vec{n}} $ and $ \tilde{g}^{\dag }_{\vec{n}} $ are
the annihilation and creation operators for the quasiparticles with spherical 
quantum numbers $\vec{n}={n,l,m}$. The wave functions $u_{\vec{n}}({\vec r})$ and 
$v_{\vec{n}}({\vec r})$ obey the standard orthonormalization rules \cite{Stringari}. 
Linearizing in the $u_{\vec{n}}({\vec r})$ and $v_{\vec{n}}({\vec r})$ 
wavefunctions, one obtains the corresponding Bogoliubov equations
\begin{eqnarray}
&&(\mathcal{H}_{GP}+U_{gg}N_{0}Q\psi _{0}^{2}Q)u_{\vec{n}}+
U_{gg}N_{0}Q\psi _{0}^{2}Q^{*}v_{\vec{n}}= 
\hbar \tilde{\omega }^{g}_{\vec{n}}u_{\vec{n}} \label{BdG1}\\
&&-U_{gg}N_{0}Q^{*}\psi _{0}(\vec{r})^{2}Qu_{\vec{n}}
-(\mathcal{H}_{GP}+U_{gg}N_{0}Q\psi _{0}^{2}Q)^{*}v_{\vec{n}}=
\hbar \tilde{\omega }^{g}_{\vec{n}}v_{\vec{n}} \label{BdG2}
\end{eqnarray}
where we have used the projection operators $ Q=1-|\psi _{0}\rangle \langle \psi _{0}| $ ($Q^{*}$)
orthogonally to $ \psi _{0} $ ($ \psi _{0}^{*} $) \cite{Castin98}, and 
$\tilde{\omega }^{g}_{\vec{n}}$ denotes the quasiparticle excitation spectrum. 

The physics of the atoms in the $|e\rangle$ state is described by the Schr\"odinger equation: 
\begin{equation}
\mathcal{H}_{e}\psi ^{e}_{\vec{m}}(\vec{r})=
\left( -\frac{\hbar ^{2}}{2m}\nabla^2 +\frac{m}{2}\omega_e^2 r^2+
2U_{ge}N_{0}|\psi _{0}|^{2}\right) \psi ^{e}_{\vec{m}}(\vec{r})=
\hbar \tilde{\omega }^{e}_{\vec{m}}\psi _{\vec{m}}^{e}(\vec{r}),
\label{excit}
\end{equation}
where we have taken into account that the $|e\rangle$ atoms 
are affected by the mean-field potential induced by the collisions with the $|g\rangle$ atoms, 
which are characterized by a coupling constant $U_{ge}=4\pi\hbar^2a_{ge}/m$, with $a_{ge}$ the 
corresponding scattering length. Due to the low density in the $|e\rangle$ trap, we have neglected 
in eq.\ (\ref{excit}) the $e$-$e$ collisions. In the following, 
$\tilde{e}^{\dag }_{\vec{m}}$ and $\tilde{e}_{\vec{m}}$ denote 
the creation and annihilation operators in the eigenstate $\psi ^{e}_{\vec{m}}$, with 
eigenfrequency $\tilde{\omega }^{e}_{\vec{m}}$.

The interaction of the atoms with the vacuum electromagnetic field is given by
\begin{equation}
\mathcal{H}_{af}=i\sum _{\vec{m}}\sum _{v}\int d\vec{k}\rho (\vec{k})
(\vec{d}\vec{\epsilon }_{kv}) \left( \eta _{0\vec{m}}^{c}\tilde{g}_{0}^{\dag }e_{\vec{m}}+
\sum _{\vec{n}}\eta _{\vec{n}\vec{m}}^{u}\tilde{g}_{\vec{n}}^{\dag }e_{\vec{m}}-
\sum _{\vec{n}}\eta _{\vec{n}\vec{m}}^{v}\tilde{g}_{\vec{n}}e_{\vec{m}}\right) a_{\vec{k}v}^{\dag }+\hc,
\label{Haf} 
\end{equation}
where $ a^{\dag }_{\vec{k}\nu } $ ($ a_{\vec{k}\nu } $) are creation (annihilation)
operators of photons of wavenumber $\vec k$, polarization $\nu$, and  
frequency $\omega_{\nu}(\vec{k})$. In eq.~(\ref{Haf}), $\vec d$ is the atomic dipole, 
$\vec\epsilon_{kv}$ is the polarization vector, and $\rho (\vec{k})$ the density of states. 
The Frank-Condon (FC) factors 
$\eta_{0\vec{m}} ^{c}=\int d\vec r \psi_{\vec{m}}^{e}(\vec{r}) \exp(i\vec k\vec r)
\psi_{0}(\vec r)$ 
characterize the transitions into the condensate, whereas  
$\eta_{\vec{n}\vec{m}}^{u}=\int d\vec r \psi_{\vec{m}}^{e}(\vec{r}) \exp(i\vec k\vec r)
u_{\vec{n}}(\vec{r})$ determine the transition into the $u$ quasiparticle wavefunctions. 
In the same way, the coefficients $\eta_{\vec{n}\vec{m}}^{v}$ 
characterize the transition into the $v$ wavefunctions.
Finally, the vacuum energy is provided by the Hamiltonian: 
$\mathcal{H}_{f}=\sum _{\nu }\int \vec{d}k\, \omega _{\nu }(\vec{k})\, 
a^{\dag }_{\vec{k}\nu }a_{\vec{k}\nu }$. 
The physics of the system is therefore described by the Hamiltonian
$\mathcal{H}=\mathcal{H}_{a}+\mathcal{H}_{af}+\mathcal{H}_{f}$,
where the atomic Hamiltonian can be written in the form (extracting the condensate energy)
$
\mathcal{H}_{a}=
\sum _{\vec{n}}\tilde{\omega }_{\vec{n}}\tilde{g}^{\dag }_{\vec{n}}\tilde{g}_{\vec{n}}+
\sum _{\vec{m}}(\omega _{0}+\tilde{\omega }_{\vec{m}}^{e})\tilde{e}^{\dag }_{\vec{m}}\tilde{e}_{\vec{m}}
$
where $\omega_0$ is the transition frequency between $|g\rangle$ and $|e\rangle$.

From the previous Hamiltonian, and using standard techniques of quantum stochastic processes 
\cite{Gardinerbooks}, we get in Born-Markov approximation the  
QME for the density matrix $\rho$:
\begin{equation}
\dot{\rho }(t)=-iH_{eff}\rho +i\rho H_{eff}^{\dag }+\mathcal{J}\rho,
\label{ME}
\end{equation}
where
$H_{eff}$
is the effective (non-Hermitian) Hamiltonian, and 
\begin{equation}
\mathcal{J}\rho = \gamma \left \{ 
\sum _{\vec{m}}\Re^{c}_{\vec{m}00\vec{m}}
\tilde{g}_{0}^{\dag }\tilde{e}_{\vec{m}}\rho \tilde{g}_{0}\tilde{e}_{\vec{m}}^{\dag }
+\sum _{\vec{m},\vec{n}} 
\Re ^{u}_{\vec{m}\vec{n}\vec{n}\vec{m}}
\tilde{g}_{\vec{n}}^{\dag }\tilde{e}_{\vec{m}}\rho \tilde{g}_{\vec{n}}\tilde{e}_{\vec{m}}^{\dag }
+\sum _{\vec{m},\vec{n}} \Re ^{v}_{\vec{m}\vec{n}\vec{n}\vec{m}}
\tilde{g}_{\vec{n}}\tilde{e}_{\vec{m}}\rho \tilde{g}^{\dag }_{\vec{n}}\tilde{e}_{\vec{m}}^{\dag }
\right \},
\label{jump}
\end{equation}
is the jump super operator, where  $\Re^{a}_{\vec{m}\vec{n}\vec{n}'\vec{m}'(\omega _{0})}
=\int d\Omega W(\Omega )\eta ^{a}_{\vec{m}\vec{n}}
(\omega _{0}\hat{e}_{k})\eta _{\vec{m}'\vec{n}'}^{a}(\omega _{0}\hat{e}_{k})^{*}$  ($a$=$c$, $u$, $v$), 
with $W(\Omega)$ denoting the emission pattern, and $\gamma$ is the spontaneous emission rate.  

The effective Hamiltonian $H_{eff}$ is related with photon reabsorption processes, whereas the jump 
superoperator describes a spontaneous emission act without any further reabsorption. In this paper, we 
consider the reabsorption processes as negligible. As commented above, this should be the case 
if the Festina Lente condition is fulfilled \cite{Castin}, or alternatively for atoms 
with a three-level $\Lambda$ scheme under the conditions of ref.\ \cite{Pfau}.  
Therefore, we constrain
ourselves to the analysis of the jump processes from $|e\rangle$ to $|g\rangle$. 
From eq.~(\ref{jump}) the amplitude for a jump from a state $\vec{n}_e$ of the $|e\rangle$ trap to 
a state $\vec{n}_g^{(a)}$ ($a=u,v,c$) of the $|g\rangle$ one is given by    
\begin{equation}
\Gamma_{\vec{n}_g^{(a)},\vec{n}_e}=
\gamma \int d\Omega W(\Omega ) |\langle \vec{n}_g^{(a)}|e^{i\vec{k}\vec{r}}|\vec{n}_e\rangle |^{2},
\label{gam1}
\end{equation}
with $\langle \vec{r}|\vec{n}_{e}\rangle =R_{l_en_e}(r)Y_{lm}(\Omega)$, and 
$\langle \vec{r}|\vec{n}_{g}^{(a)}\rangle =R_{l_gn_g}^{(a)}(r)Y_{lm}(\Omega)$.
The FC factors are of the form
\begin{eqnarray}
\langle \vec{n}_g^{(a)}|e^{i\vec{k}\vec{r}}|\vec{n}_{e}\rangle = 
4\pi \sum ^{\infty }_{l''=0}\sum ^{l''}_{m=-l''}(i)^{l''}Y^{*}_{l''m}(\Omega _{k})
 A^{l''(a)}_{n_gl_gn_el_e}B^{l''}_{l_{g}m_{g}l_{e}m_{e}}, 
\label{FK}
\end{eqnarray}
where $A^{l''(a)}_{n_gl_gn_el_e}=\int drr^{2}\; R^{a}_{l_gn_g}(r)\; j_{l''}(kr)R_{l_en_e}(r)$ is the radial 
integral. The angular part is provided by $B^{l''}_{l_{g}m_{g}l_{e}m_{e}}=\int d\Omega _{r}Y^{*}_{l_{g}m_{g}}(\Omega _{r})
Y_{l''m}(\Omega _{r})Y_{l_{e}m_{e}}(\Omega _{r})$, 
solvable in terms of the corresponding 3-j symbols. Due to the properties of these symbols, 
$B^{l''}_{l_{g}m_{g}l_{e}m_{e}}$ is independent of $m$ since  $-m_{g}+m_{e}+m=0$, $ l_{g}+l_{e}+l''=2p $
(where $p$ is an integer), and $ |l_{g}-l_{e}'|\leq l''\leq l_{g}+l_{e} $. From the amplitudes (\ref{gam1}) and 
the populations of the corresponding levels, we obtain the transition probabilities
describing creation of a quasiparticle $n=(n_g,l_g)$
\begin{equation}
P^{u}_{n}  =  (\langle N_{n}(T)\rangle +1)\sum _{n_el_e}
\langle N_{\vec {n}_e}(T_{e})\rangle \sum _{m_g,m_e} \Gamma _{\vec{n}_g^{(u)},\vec{n}_e}. 
\label{Pu}
\end{equation}
Similar expressions are obtained for the transition probability of destruction of a quasiparticle 
$n$ associated with the transfer of a particle into the condensate after thermalization, 
$P^{v}_{n}$, and the direct decay process into the condensate, $P^{c}$. In eq.~(\ref{Pu}), 
$ \langle N_{n} (T)\rangle =(\exp [E_{n}/K_{B}T]-1)^{-1} $
is the Bose-Einstein distribution in the $|g\rangle$ trap, and 
$\langle N_{\vec{n}_e} (T_{e})\rangle =\hat{N}\exp [-E_{\vec{n}_e}/K_{B}T_{e}]$
is the Boltzmann distribution (normalized to $ 1 $) in the $|e\rangle$ trap.

Knowing the transition probabilities $P^{u}_{n}$ (\ref{Pu}), $P^{v}_{n}$ and $P^{c}$
we can simulate the effects of a spontaneous emission process. 
We assume that the time scale of thermalization in the $|g\rangle$ trap is much faster 
than the time interval between different 
pump events, {\it i.e.} the collisional rate in the ground-state trap is much larger than $\gamma$. In this way, after 
every pump, the system re-thermalizes, redistributing the energy gained or lost during the pump process. 
We analyze in the following the variation of the condensate temperature after the pumping. 

After pumping, the number of $|g\rangle$ atoms increases to $N'=N+1$, 
and the energy becomes:
\begin{equation}
\label{EP}
\left\langle E' \right\rangle =E+\mu P^{c}+
\sum _{n}(\epsilon _{n}+\mu C_{n} (P^{u}_{n}-P_{n}^{v}).
\end{equation}
where $E$ is the energy before the pumping, and 
$C_{n}=\left| u_{n}\right| ^{2}+\left| v_{n}\right| ^{2} $.
 After the re-thermalization, 
the system acquires a new temperature $T'$ which can be evaluated from the expressions:
\begin{eqnarray}
E'& = & \langle \psi_{0} | {\cal H}_{G} - \frac{U_{gg}}{2} \left|\psi_{0}\right|^{2}   |\psi_{0} \rangle  
        +\sum _{n}N_{n}(T')(\epsilon _{n}
        + \mu C_{n}) 
	- \sum _{n}(\epsilon _{n}-\mu )|v_{n}|^{2}, \label{Ee}\\
N_{0}(T') & = & N+1-\sum _{n}N_{n}(T')C_{n}+|v_{n}|^{2} \label{N0p},
\end{eqnarray}
where $N_{n}(T)=(\exp [\epsilon _{n}/K_{B}T]-1)^{-1}$ is the Bose-Einstein distribution in 
the $|g\rangle$ trap.
In (\ref{Ee}) and (\ref{N0p}), we assume the excitation spectrum 
as that calculated for $N$ particles, {\it i.e.} before the pumping. This approximation is valid as long as the 
number of excited particles $N-N_0\gg 1$. 

Assuming $T'=T+\delta T$ with $\delta T\ll T$, we can perform a Taylor expansion up to 
first order in $\delta T/T$. Substituting into eqs.~(\ref{Ee}) and (\ref{N0p}), we can obtain the expression 
for $\xi=(N'_0/N'-N_0/N)/(N_0/N)$. If $\xi>0$, the pumping increases the 
relative population in the condensate, and therefore the temperature of the sample is reduced. 
On the contrary, if $\xi<0$ the system is heated by the pumping. It is easy to see that up to terms 
of the order $\delta N/N_0^2$, where $\delta N=N-N_0$, the system is cooled ($\xi>0$) when the condition:
\begin{equation}
(N_{0}-1)\frac{\tilde{\mu }(N_{0})(1-P^{c})-\sum _{n}(\tilde{\epsilon }_{n}+
\tilde{\mu }C_{n})(P^{u}_{n}-P^{v}_{n})}{\sum _{n}\tilde{\epsilon }_{n}^{2}
\frac{\exp [\tilde\epsilon _{n}]}{(\exp [\tilde\epsilon _{n}]-1)^{2}}}
\sum _{n}\tilde{\epsilon }_{n} C_{n}
\frac{\exp [\tilde\epsilon _{n}]}{(\exp [\tilde\epsilon _{n}]-1)^{2}}  >\delta N
\label{cond}
\end{equation}
is fulfilled, with $\tilde{\mu }=\mu /K_{B}T$ , $\tilde{\epsilon_{n}}=\epsilon_{n} /K_{B}T$.  
For moderately low $T$ this condition is typically fulfilled.
For large $T$, $\delta N$ can increase enough to violate (\ref{cond}). 
On the other hand for very low $T$ the transition probability into the condensate 
$P^{c}\rightarrow 1$, and also in this case the condensate is heated. Hence, there 
is a transition from heating to cooling at some finite threshold temperature,  
$T_{th}$, which depends on four different parameters: the total number of atoms $N$, 
the temperature, $T_{e}$, of the excited state trap, the coupling constant $U_{eg}$ for the collisions between 
$|e\rangle$ and $|g\rangle$ atoms, and the ratio $\omega _{e}/\omega _{g}$
between the frequencies of the $|g\rangle$ and $|e\rangle$ traps.
When $N$ increases, so does $\tilde\mu$ and it is easy to see that $T_{th}$ decreases.
If $ T_{e} $ increases the transition to higher lying excited states
are more favorable, and as one could expect $T_{th}$ increases.
The parameters $ U_{eg} $ and $ \omega _{e}/\omega _{g} $ influence 
the spectrum of the excited state trap. For $U_{eg}=0$ the harmonic oscillator 
spectrum is recovered. However, when $U_{eg}$ grows, the mean-field repulsion leads to 
an expulsion of the $|e\rangle$ atoms from the trap center, increasing the probability for a 
transition into excited states of the $|g\rangle$ trap. 
On the other hand, if $\omega_e\gg\omega_g$, and $T_e$ is sufficiently small, the mean-field 
shift just provides an overall homogeneous shift of the levels of the $|e\rangle$ trap, and the 
$T_{th}$ becomes comparable to that for $ U_{eg}=0 $.

We have analyzed $\delta T$ after a pumping using the following procedure. 
For given values of $N$ and $T$, we first 
calculate the number of atoms in the BEC, $N_0$, and evaluate $\psi_0$ by evolving eq.~(\ref{GPE}) 
in imaginary time. Then, we diagonalize eqs.~(\ref{BdG1}) and (\ref{BdG2}) employing 
a harmonic oscillator basis, to find the $\{ u,v\}$ eigenfunctions, and the corresponding eigenenergies
\footnote{
In TF regime $u_n$,$v_n$ and $\epsilon_n$ are known analytically. We employ here however the numerical
          approach, which works in all intermediate regime between weakly condensed and TF regime}
. As a next step, we calculate $\Gamma _{\vec{n}_{g}^{(a)},\vec{n}_{e}}$ ($a=c,u,v$) using eq.~(\ref{gam1}). For a given 
$T_e$ we evaluate 
$P^{u}_{nl}$, $P^{v}_{nl}$ and $P^{c}$. 
From eq.~(\ref{EP}) we obtain the new average energy after 
the pumping. Finally, from eqs.~(\ref{Ee}) and (\ref{N0p}) we obtain the new temperature after the rethermalization.  

\begin{figure}[t]
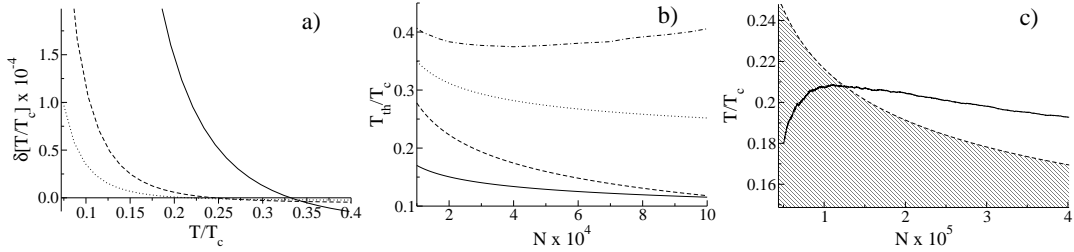

\threeimages[scale=0.18]{fig1.eps}{fig2.eps}{fig3.eps}
\caption{a) variation of $T/T_c$ after a spontaneous emission as a 
            function of the initial $T/T_c$
            for $\omega=20\times 2\pi$ Hz, $\omega_e=10\omega_g$, 
       	    $T_e=21 \hbar \omega_g$, 
            and $N=10^4$, $5\times 10^4$ and $10^5$;
         b) dependence of $T_{th}$ on $N$ 
	    for $\omega_e=\omega_g$   
	    and $U_{eg}/U_{gg}=1$(dotted--dashed curve), $0.5$ (dotted curve), 
	    and $0$ (dashed curve); the solid curve shows the case 
	    $\omega_e=10 \omega_g$ and $U_{eg}/U_{gg}=1$; 
         c) solid line indicates the Monte Carlo simulation of the loading process and dashed line $T_{th}$ for 
	    $\omega_e=10\omega_g$, $T_e=21 \hbar \omega_g$ }
\label{fig:1}  
\end{figure}

Fig. ~\ref{fig:1}a shows the variation of $T/T_c$ after a spontaneous emission as a function of the initial $T/T_c$
for $\omega=20\times 2\pi$ Hz, $\omega_e=10\omega_g$, $T_e=21 \hbar \omega_g$, and $N=10^4$, $5\times 10^4$ and $10^5$.
In the calculation of the FC factors (\ref{FK}), we have considered a Lamb-Dicke parameter 
$\eta^2\equiv E_{rec}/\hbar\omega_g=1$, where $E_{rec}$ is the photon recoil energy. 
The choice of $\eta$ and $T_e$ is due to our numerical limitations, but we expect qualitatively similar 
results for other values of these parameters.
As previously discussed, the system is cooled for $T>T_{th}$, where $T_{th}$ is 
progressively lower for larger $N$. In fig.~\ref{fig:1}a, due to the low value of $T_e$ chosen, it is not possible 
to observe the heating region for large temperatures, which for this case is located at $T$ close to $T_c$. 
For other ranges of parameters the qualitative picture does not change significantly, {\it i.e.} always a threshold 
temperature separating heating and cooling regions exists.

We have analyzed the dependence of $T_{th}$ on different parameters. Fig. ~\ref{fig:1}b shows the 
dependence of $T_{th}$ on $N$ for $\omega_e=\omega_g$ and different values of $U_{eg}$. For $U_{eg}\rightarrow 0$, 
$T_{th}$ decays monotonically with $N$. However, when $U_{eg}$ increases, and due to the 
repulsive mean-field induced by the $e$-$g$ collisions, the eigenfunctions of the $|e\rangle$ trap possess a local minimum 
at the trap center, and the pumping into the condensate becomes less effective. Consequently, $T_{th}$ increases 
for large $N$, as observed for $U_{eg}=U_{gg}$. This effect is less pronounced, for larger values 
of $\omega_e$ keeping fixed $T_e$, since then the lower levels of the $|e\rangle$ trap have an extension smaller than 
the BEC wavefunction, and hence the mean-field just produces a global energy shift of the lowest 
$|e\rangle$ levels, with no consequences for the pumping (see fig. ~\ref{fig:1}b).

Finally, we have simulated for different conditions the continuous optical pumping into the BEC. 
To this aim, we have employed Monte Carlo methods to evaluate the corresponding rate equations, with 
transition probabilities $P^{u}_{nl}$ (\ref{Pu}), $P^{v}_{nl}$ and $P^{c}$.
In principle, after every pumping the previously described 
algorithm should be repeated. In practice, it is enough to do so after a number of pumps 
$N_{pumps}\ll \delta N$, which is dynamically adjusted in our calculations. Fig. ~\ref{fig:1}c shows our 
results for $\omega_e=10\omega_g$, $T_e=21 \hbar \omega_g$.  
The dashed curve in fig.~\ref{fig:1}c indicates $T_{th}$. 
The simulation is started within the heating region (below the $T_{th}$ curve), 
with $N=4.8\times 10^4$, and $T/T_c=0.18$. As expected, $T$  
increases with the pumping. However, once $(N,T/T_c)$ crosses the $T_{th}$ curve, the temperature begins to decrease 
when pumping. Therefore, interestingly, the temperature of the system should asymptotically approach the curve 
$T_{th}$. 

In this paper, we have analyzed the continuous optical loading of a BEC in the Thomas-Fermi regime. 
Contrary to the weak-condensation case, the BEC and excited-state 
wavefunctions vary during the loading. 
By means of GPE and Bogoliubov equations, we have determined before every pumping 
the proper wavefunctions, and transition probabilities. 
Assuming rapid thermalization, we have monitored the variation of the 
temperature during the loading.
We have observed that for a given number of trapped atoms, $N$, there is always a threshold temperature, $T_{th}$, 
below which the BEC is heated. 
The concept of $T_{th}$ is very general and should also apply to collisional loading of a BEC.
We have analyzed the dependence of $T_{th}$ on different parameters, 
in particular $N$, the trap geometries, and the interparticle interactions. A lower $T_{th}$ is obtained 
for smaller interactions between excited-state and ground-state atoms, since repulsive interactions 
displace the excited-state atoms away from the trap center, preventing the effective pumping into the BEC.
Our analysis shows that, not only the number of atoms in the trap, but also the temperature of the system, can be 
maintained by means of optical loading. To this aim, the condensate should be created with a temperature and a 
number of atoms, lying close to the curve $T_{th}(N)$. In that case, in the presence of losses, 
if the loss mechanism leads to heating, $T$ becomes larger than $T_{th}(N)$, and the system is cooled 
by next pumpings, whereas the opposite is true if the loss mechanism cools. Interestingly, this could allow 
for the sustained analysis of BEC at a quasi-constant temperature.

Several physical scenarios could be devised for the continuous optical loading of a BEC. For example, 
an optical lattice could be used to move atoms in an internal state $|r\rangle$ 
from a relatively hot reservoir into the center of the BEC region, where a Raman pulse 
could be employed to transfer the $|r\rangle$ atoms into an state $|e\rangle$, which decays into the ground-state $|g\rangle$. 
In this way, it could be achieved that only those lattice sites near the condensate center actually decay into the $|g\rangle$ 
trap. For a sufficiently strong optical lattice each lattice site will behave as a single trap with few 
occupied levels, as that discussed in our paper. Interestingly, in this scenario, due to the small size of the $|e\rangle$ trap, 
$T_{th}$ should monotonically decay with $N$, since the mean-field effects should just shift globally the states of the $|e\rangle$ trap. 
Therefore, in this system the pumping could be employed to post-cool an already formed BEC down to $T=0$, and maintain the 
condensate temperature very low against possible loss sources.

We acknowledge support from the Alexander von Humboldt Stiftung, the Deutsche 
Forschungsgemeinschaft, EU RTN Network "Cold Quantum Gases", ESF PESC Program BEC2000+.

\end{document}